# Improved visualisation of patient-specific heart structure using three-dimensional printing coupled with image-processing techniques inspired by astrophysical methods

**I. Brewis and J.A. McLaughlin**

Northumbria University, Newcastle upon Tyne, NE1 8ST, United Kingdom

**Abstract**
The aim of this study is to use image-processing techniques developed in the field of astrophysics as inspiration for a novel approach to the three-dimensional (3D) imaging of periprocedural medical data, with the intention of providing improved visualisation of patient-specific heart structure and thereby allowing for an improved quality of procedural planning with regards to individualized cardiovascular healthcare. Using anonymized patient DICOM data for a cardiac computed tomography (CT) angiography, two-dimensional slices of the patient's heart were processed using a series of software packages in order to produce an accurate 3D representation of the patient's heart tissue as a computer-generated stereolithography (STL) file, followed by the creation of a tactile 3D printout. We find that the models produced provide clear definition of heart structure, in particular in the left atrium, left ventricle and aorta. This level of clarity also allows for the aortic valve to be observed and 3D printed. This study provides a step-by-step blueprint of how this can be achieved using open source software, specifically Slicer 4.8.1, MeshLab and AutoDesk Netfabb. In addition, the implementation of astrophysical image-processing techniques shows an improvement in modelling of the heart based on the CT data, in particular in the case of small-scale features where echocardiography has previously been required for more reliable results.

# Introduction

## The Merits of 3D Printing

In recent years, due to the increasing complexity of cardiovascular conditions, surgical and interventional procedures have called for an increase in advanced periprocedural imaging and have called for the production of tactile models in order to provide an expedient and safe approach towards dealing with these complex conditions[1]. Using either magnetic resonance imaging (MRI), computed tomography (CT), three-dimensional (3D) echocardiography or a combination of the three, it is now possible to produce a patient-specific 3D rendering of the heart[1]. These 3D renderings can provide a far clearer picture to surgeons and patients alike as to the problems which may be encountered during surgery when compared with conventional 2D images. One example of such advantages is the ability to view a 3D replica of damage to heart tissue and structure from multiple angles as opposed to typical two-dimensional (2D) scans which are often limited to up to three views of the patient's anatomy; the axial (top down), sagittal (side on) and coronal (front facing) views. In addition to this, 3D renderings, when printed, produce a tactile model which can be used to observe a range of heart conditions from valve deformation in congenital heart disease[1] to tissue scarring resulting in arythmmia[2]. These models can then be utilized as a more risk-free alternative to acquainting new or training surgeons as to the various cardiovascular afflictions as well as how best to treat them. In addition to this, 3D models can be utilized as a tool for consulting doctors, providing patients with a better understanding as to their condition[2] and as an explanatory tool to show what will occur during their planned procedures. 3D modelling of medical imaging data also allows for more accurate transcatheter aortic valve interventions[1] (TAVI) since 3D models provide a risk free environment to simulate procedures such as this without any dependence on digital storage media or the need for a computer[1].

Olejník et al[3] defends the accuracy of 3D renderings based upon medical imaging data, showing in their 2017 paper that current 3D models can show as little variation between actual and rendered heart feature dimensions as $0.18 \pm 0.38$mm thereby providing strong incentive towards the use of tactile 3D models and their three-dimensionally rendered computer-image counterparts as both an informative and educational tool for patients and surgeons alike.

## The Need for Further Medical Image-Processing Research

Whilst current 3D renderings are observed to be relatively accurate, there is still an observed margin of error between 3D printed models and patient physiology[3]. In addition, image processing of patient data can be time consuming and costly. This paper explores whether the inclusion of image-processing techniques common in astrophysics can be applied to potentially further reduce this margin for error between patient and model, and the possible inclusion of astrophysical image-processing algorithms could reduce image-processing timescales and provide a more rapid transition from DICOM image data to 3D volume rendering.

## Astrophysical Impact

Astrophysical image-processing techniques have seen an increase in automated image processing techniques[4]. Developments within the field of image processing have led to several techniques which prove instrumental in the development of 3D renderings based off of CT image data. These image-processing techniques include *vignetting*[4], *boxcar smoothing functions*[4], *dilation*[5], *edge detection*[5] and the development of a pixel platescale such as is commonly used for charge-coupled devices (CCDs) such as on-board the SOHO[4,6], TRACE[4,7] and STEREO[4,8] spacecraft where the occurrence of "dark" and "hot" pixels which can lead to false detections of small-scale phenomena if not properly treated during the data-processing stages[4].

This paper aims to provide a step-by-step blueprint of the methodology for producing a volume rendering of a patient's heart starting with the DICOM data and ending with the creation of a tactile 3D printout. This paper describes in detail how this can be achieved, specifically using (free) open source software, namely *Slicer 4.8.1*, *MeshLab* and *AutoDesk Netfabb*.

# Methodology

The methodology for producing a volume rendering of the patient's heart consisted primarily of five main procedural steps:
1. Data Acquisition
2. Segmentation
3. File Conversion
4. File Fixing and Design
5. 3D Printing

## 1. Data Acquisition

By working in partnership with the Freeman hospital in Newcastle upon Tyne, UK, it was possible to acquire the CT scan data for a single patient who, in the interest of safety and ethics, was made anonymous during this research. The CT scan data for the patient was captured in the standard DICOM image file format. A series of 856 CT images of the patient's chest were provided for the axial, coronal and sagittal directions of the scan in order to provide a three-dimensional series of slices taken 1mm apart from each other to be later processed and rendered. These images however could not be utilized directly in their DICOM file format by 3D printers, and therefore had to be converted into another format which a 3D printer would recognise, namely a stereolithography (STL) file. This format defines the surface of the model as being composed of a series of interlinking triangles[1,9].

## 2. Segmentation

The segmentation process allows for the various structures observed within the DICOM image files to be identified, labelled and reconstructed as a three-dimensional model of the extracted feature(s). The region-of-interest (ROI) in this case was selected as the heart tissue.

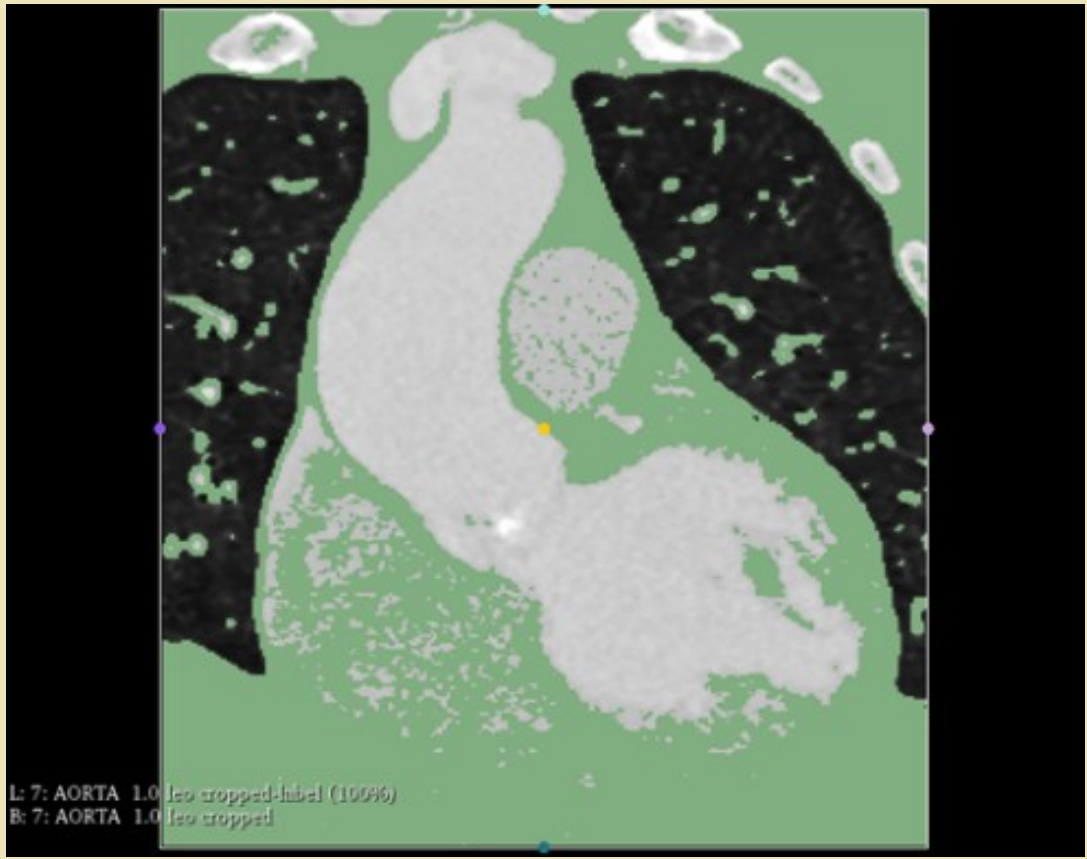

**Figure 1** Highlighted data points (green) obtained from initial thresholding. The colored dots represent the borders of the segmented ROI with the yellow dot representing the centre.

Using the open source software package Slicer 4.8.1, the patient's DICOM image data was loaded, allowing for the axial, sagittal and coronal views of the patient's chest cavity to be observed simultaneously in 2D.

As an initial step, a ROI was established around the heart allowing for excess data to be cropped from the DICOM image data. The remaining image data then gave a clearer view of the heart for use during the segmentation stage.

## 2.1 Full Heart Segmentation

Initial feature extraction required that all heart tissue should be incorporated into the 3D model whilst minimizing the amount of excess data such as from the blood or bone so as to provide an accurate representation of both the internal and external heart structure. Due to the contrast in the pixel platescale[4] between the heart tissue (dark grey) and areas of blood (light grey) and bone (white), areas of bright or "hot" pixels could be removed through the use of thresholding[5].

Through a similar process, areas of "dark" pixels such as regions of lung could also be removed through thresholding. Thresholding is defined as the process of applying a limited pixel platescale to a data set in order to remove areas where pixels are either too bright or too dark. This is carried out by assigning a range of numerical intensity values to each pixel observed where darker areas are given lower numerical values (low intensity) and lighter areas are given higher values (high intensity). For an 8-bit display, the range of intensity, or pixel depth, allows for a range of 256 different intensities (i.e. shades of grey) to be stored. In the case of the DICOM data provided, a 16-bit greyscale display was used. Slicer 4.8.1 would automatically adjust this scale should no pixels outside of the whole image platescale be observed, typically giving a pixel depth of [-600, 1055] for the data provided. By applying a threshold to the image data, only pixels lying within the same platescale as the heart tissue were included. Slicer 4.8.1 software comes with an in-built threshold effect allowing for all pixels within a specified platescale within the ROI to be highlighted. Using the threshold effect tool in

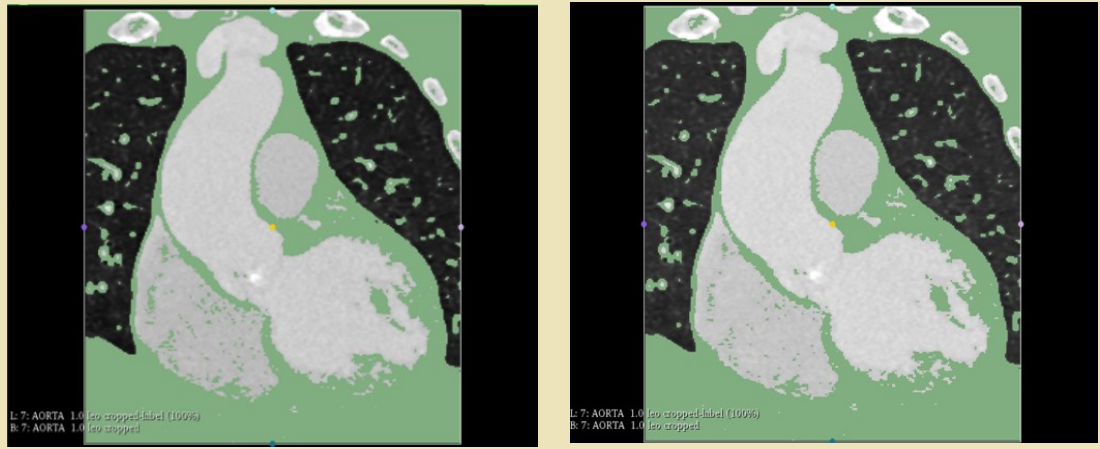

**Figure 2** Segmented heart data following first thresholding of right heart region (left) and segmented heart data following second thresholding of right heart region (right). The right ventricle (bottom left) shows improved clarity of internal chamber segmentation following second thresholding.

Slicer 4.8.1, the optimum threshold range to ensure the greatest amount of heart tissue obtained was found to lie in the range [-600, 130]. This can be seen in Figure 1.

Initial thresholding highlighted all heart tissue, with the inclusion of deoxygenated blood in both the right atrium and right ventricle. Areas of deoxygenated blood were found to be removed by applying a second threshold in the range [-600, 70] to the chambers of the right heart. It should be stated that the application of this second threshold to the entire ROI would have resulted in loss of regions of heart tissue, thus sacrificing clarity of the model. It was therefore more practical to apply the new threshold to the right heart regions alone using Slicer 4.8.1's in-built eraser and draw tools in order to select a specified region slice-by-slice and remove data lying inside the specified threshold.

In order to minimize human error, an initial threshold for the eraser tool was selected at [80, 130] to allow for minimal loss of heart tissue around the edges of the right heart chambers whilst still removing the bulk of the unwanted blood, allowing for the internal tissue structure of these chambers to be observed. In order to allow for a greater degree of precision during this, a zoomed in view of the coronal slice alone was selected and each subsequent slice processed individually.

Once applied to all slices, an additional pass was made across each slice using the draw and erase tool for a now lower threshold of [70,130] in order to further reduce the number of unwanted or "bad" pixels[4] observed within the right heart chambers. This can be seen in Figure 2. Once the second pass of thresholding was completed, the thresholded heart data was then converted from a series of segmented 2D slices into a 3D rendering of the data using Slicer 4.8.1's model maker tool. Once a 3D rendering was completed, the model was then converted to .stl format for further processing and saved under the file name Model_1.stl.

This method, whilst modelling the majority of the heart's internal structure, omitted small-scale features such as valves. In order to extract these small-scale features from the DICOM data available, an alternative approach was introduced.

## 2.2 Aortal Valve Segmentation

By highlighting the aortal region of the heart in the neighborhood of the aortal valve, and using *vignetting*[4] to remove regions of brighter pixels from the borders of the images around the valve region, we were able to reduce background interference. This resulted in a clearer picture of the valve structure, since the darker background tissue was largely removed[5,10].

Next, a higher threshold of [-600, 170] was applied in order to incorporate the brighter pixel platescale of the valve tissue into the model. Due to a large amount of calcification observed on the valve as a white region, i.e. a large build-up of calcium present on the aortic valve cusps, additional (second) thresholding was applied to the calcified region using Slicer 4.8.1's draw tool for a second threshold of [320, 1055].

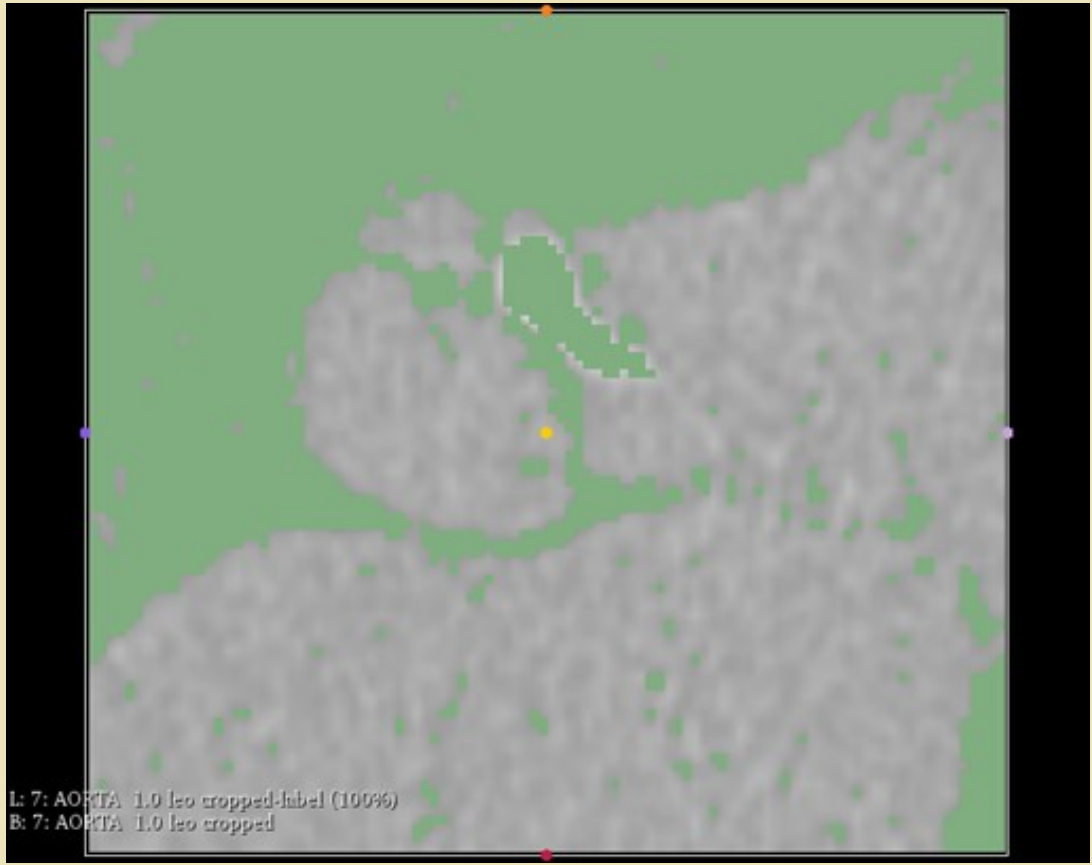

**Figure 3** Segmented aortal valve region with included calcified region.

With the calcified region then highlighted to be included in the 3D rendering, it was then possible to observe the three aortic valve cusps and the tricuspid pattern where individual valve cusps partially overlap and occlude each other[11]. Using an axial view of the segmented valve, further enhancement of the tricuspid pattern was applied using Slicer 4.8.1's paint tool for a spherical brush of radius 1.400mm along the tricuspid lines of occlusion for a threshold range of [170, 190] between the axial slices -977.768mm to -957.270mm. This can be seen in Figure 3.

The result of this processing stage then allowed for areas of aortic valve cusp occlusion and areas of bad pixels to be distinguished by the clarity of the image observed.

Using the occlusion lines observed between axial slices -977.768mm to -957.270mm as a mask[11], thresholding using Slicer 4.8.1's draw tool was selected for a threshold range [130, 190] around areas where valve cusp outlines could be observed. In order to minimize error, and to provide a higher degree of clarity to the observed valve tissue, additional thresholding passes were applied to axial slices -977.768mm to -957.270mm for incrementally increasing thresholds of [190, 200], [200, 210] and [210, 220]. This can be seen in Figure 4.

Following thresholding, areas of unwanted segmented blood were removed using Slicer 4.8.1's draw and eraser tool for a threshold range [130, 220] across all coronal view slices. This allowed for the internal structure of the aorta to be observed and for the valve to not be obscured by excess data, thus producing the desired "hollow heart" model. This change-of-view also allowed for falsely-segmented areas from the axial view thresholding steps to be removed producing a clear picture of the tricuspid valve pattern in three-dimensions.

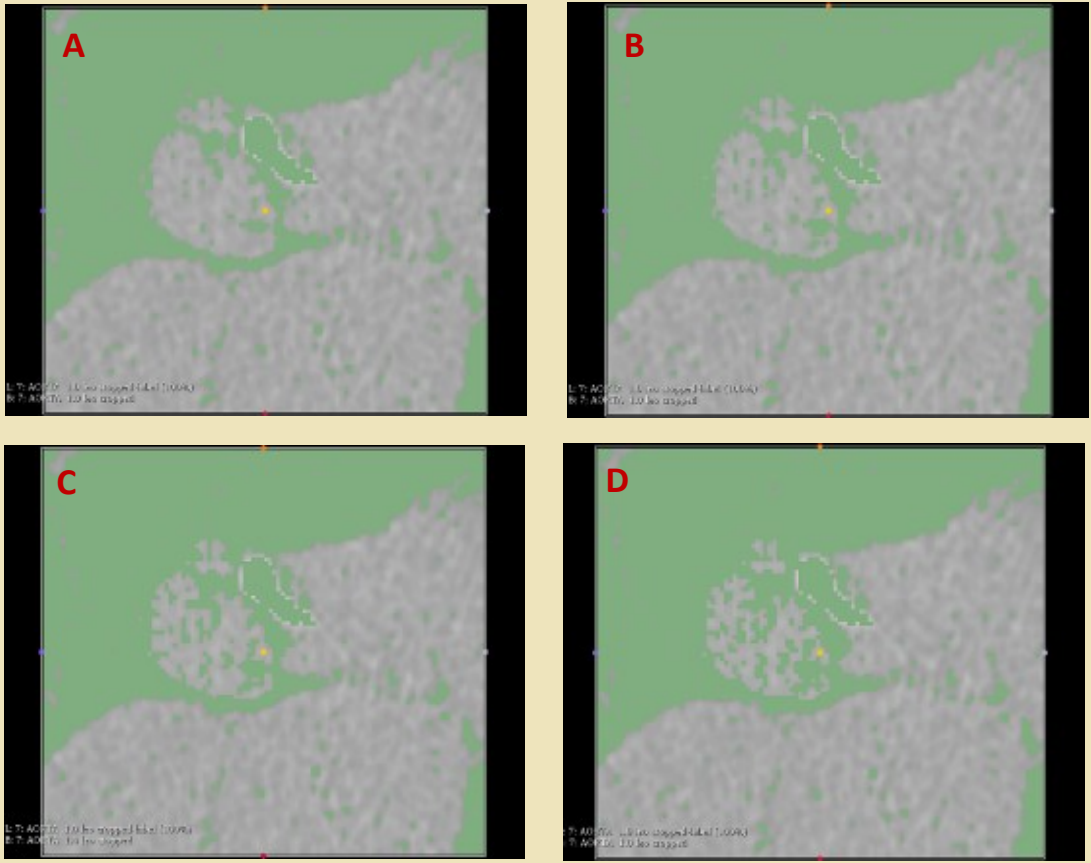

**Figure 4** Incremental threshold passes from axial slice viewer. Progressing from A to D shows passes [130, 190], [190, 200], [200, 210] and [210, 220].

The resulting segmented volume was then subsequently dilated[5] in preparation for thinning[5,11] during the design-fixing stage of processing. *Dilation*[5] of the image data extends any highlighted data point outward to its nearest neighbouring pixels in the axial, sagittal and coronal slices in accordance with the expression[5]

$$\underline{A} \oplus \underline{B} = \left\{ z \left[ (\hat{\underline{B}})_z \cap \underline{A} \right] \subseteq \underline{A} \right\} \qquad (1)$$

where $\underline{A} \oplus \underline{B}$ is the dilation of a data set $\underline{A}$ by structuring element $\underline{B}$ which are defined as sets in $\mathbb{Z}^2$.

For both the full heart and the aortal valve, the total time taken for image segmentation was on the order of tens of minutes, where some parts of the segmentation process were, naturally, more time-consuming than others. For the method presented here, the main time-consuming step was whilst utilizing Slicer 4.8.1's in-built eraser and draw tools.

## 3. File conversion

The resulting segmented model was then converted into a 3D rendering using the model maker function in Slicer 4.8.1 and exported in .stl format for further processing under the file name Aorta_1.stl.

## 4. File Fixing and Design

Using the software package MeshLab[12], a *boxcar smoothing function*[4,13,14] was applied to model Aorta_1.stl by taking the average intensity of each individual pixel with its nearest three neighbors in order to remove any unwanted pixels from the border edges of the valve[9] whilst maintaining the overall clarity of the valve cusps. This can be seen in Figure 5. This processing technique is traditionally used in astrophysics in order to assist with feature extraction from solar image data by removing areas of ambient light to sharpen regions such as coronal loops. For example, Thurgood et al[13] and Weberg et al[14] both utilize boxcar smoothing as a pre-processing technique in order to suppress unwanted noise from solar image data, in order to observe transverse wave behavior present within the solar wind. Weberg et al[14] goes on to state that the use of this technique reduced the negative impact of image noise in later processing steps and contributed towards the measurement of transverse wave amplitudes within 1% - 2% accuracy when compared with simulated results.

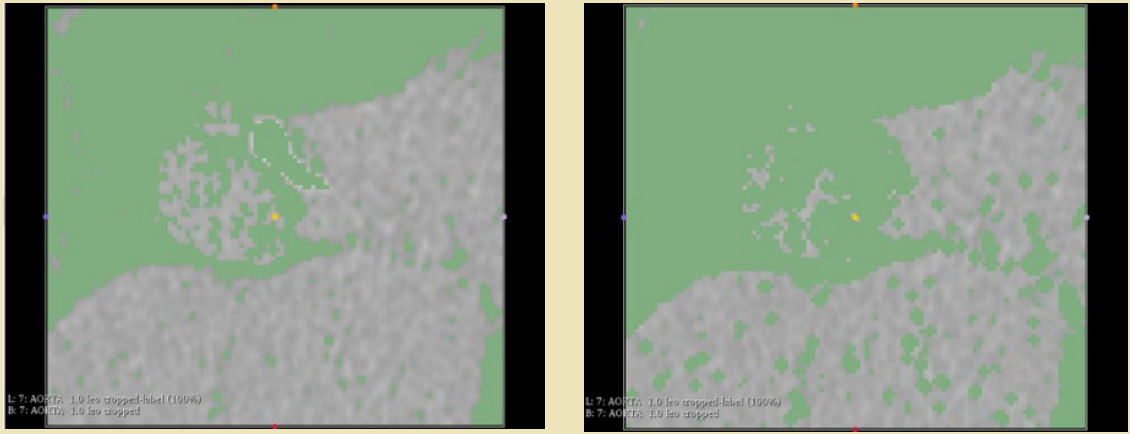

**Figure 5** Valve rendering prior to (left) and post (right)

It was realised that this technique could be used in a similar capacity within medical image processing to better highlight small-scale features such as the aortal valve cusps. Once smoothed, the new, fixed valve model was then saved and exported for further processing.

Following this stage, both the complete heart (Model_1.stl) and aortic valve (Aorta_1.stl) models were imported into the software package AutoDesk Netfabb for file fixing in preparation for 3D printing. Both volume renderings were subjected to the same pre-print processing stages within this software package. Each .stl model was imported and had its largest respective shell extracted from the main bulk of the model so as to remove any disconnected or floating points which remained independent of the rendered heart tissue. All remaining shells were then deleted leaving only the single desired extracted shell. Each model was then divided into sections, two for Aorta_1.stl and three for Model_1.stl, to reveal both the internal heart chambers and an internal view of the aortal valve.

## 5. 3D printing

Following dividing, models were saved and printed using a combination of a powder printer and a Stereolithography additive manufactured (SLA) liquid-based jetted Photopolymer printer[15].

The resulting 3D printouts were created using a powder printer for the complete heart model and an SLA printer for the aortic valve rendering. Powder printing, in general, provides accurate and free standing models which required little post-processing in order to allow them to maintain their shape, making them perfect for replicating the complex internal structure of the heart chambers and major blood vessels. The SLA printer produced more sturdy 3D models yet could not produce models which did not require additional support in order to maintain the desired structure of the internal heart chambers (i.e. free standing models) during printing. For a simple and relatively flat structure however, such as the aortic valve, the free standing issue was not a concern and the stronger resin-based model proved to be ideal for producing thin and delicate structures such as the tricuspid valve (see Figure 6).

## Results

In the case of the full 3D printed heart model (model_1.stl) pictured in Figure 7 (left), a clear 3D rendering of the internal structure of the four heart chambers can be observed as well as the major blood vessels including the aorta, superior vena cava and the pulmonary vein and artery. The patient in question whose heart had been modelled was suffering from both an aortal aneurysm, i.e. the weakening of the tissue lining the walls of the aorta, and from calcification of the aortic valve.

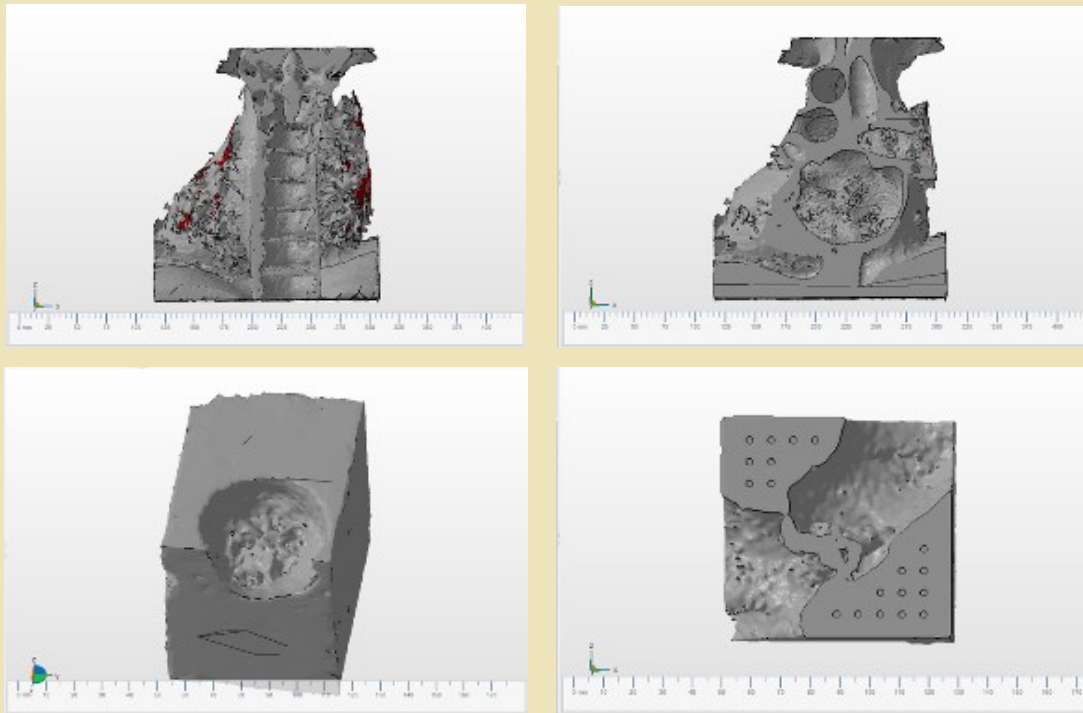

**Figure 6** Netfabb image processing stages for model_1.stl (above) and Aorta_1.stl (below). From left to right the processing stages can be observed; initial and divided models.

The aortic aneurysm can be observed clearly on the heart models as a depression within the side wall of the aorta located above the left ventricle. A clearer observation of the aortic aneurysm however can also be seen in the valve model (aorta_1.stl) pictured in Figure 7 (right), just above the aortic valve. In addition to this, aorta_1.stl also included the calcified regions present on the patient's aortic valve, allowing for the extent of the calcium deposit on the valve to be observed in three-dimensions. This level of detail and insight into patient specific anatomy has been shown previously to be instrumental in perioperative planning[16,17]. As a result of this, the use of 3D models can reduce the need for surgeons to improvise, can save on intraoperative time and improve the chance of a better outcome[3,18].

## Limitations of Study

We have successful processed DICOM image data into tactile 3D models, however the study was limited by the pre-set number of image-processing tools built into the (free) open source software Slicer 4.8.1. The software lacked the ability to program and add more complex astrophysical image-processing tools and algorithms such as *region-based feature detection*[4] and *Hough Transforms*[5]. Global processing of the greyscale DICOM image files using the Hough Transform for example would allow for each point in an image consisting of $n$ points to be compared with the greyscale intensity gradient of nearby points. Using this intensity scale, borderlines of the image could be obtained in order to extract the heart model without the inclusion of excess shells or regions of diaphragm and muscle tissue surrounding the spine, thus simplifying the processing stages. The Hough Transform in particular would provide accurate results by considering the parameter space $b = -x_i a + y_i$ for a line $y_i = ax_i + b$ in order to find the intercept within parameter space $(a', b')$ between each combination of $x_i$ and $y_i$ which could simultaneously lie on the line $y_i$.[5] In addition, the inability to filter DICOM image data prior to segmentation resulted in the appearance of regions of noise within the images themselves. The use of boxcar smoothing techniques at this earlier stage could have given the image of heart

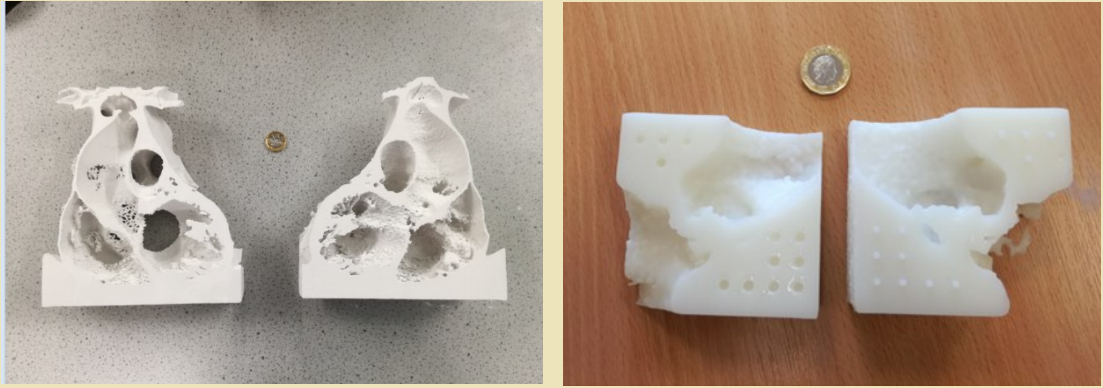

**Figure 7** 3D printed heart models. (left) complete heart model (model_1.stl), (right) aortic valve model (Aorta_1.stl). UK one pound coin for scale.

tissue in greater clarity thus improving later image-processing stages.

Another limiting factor of the software package Slicer 4.8.1 was the inability to apply region-specific threshold values. For example, in order to remove regions of both oxygenated (light grey) and deoxygenated (dark grey) blood from the volume rendering of the heart, separate threshold values were required to be applied. The inclusion of an adaptive thresholding feature[5] in order to separate the various heart chambers into sub-images for region-specific thresholding could improve image-processing time whilst minimizing both tissue loss and possible thresholding errors encountered when thresholding by hand.

Despite software limitations reducing the complexity of image-processing techniques available to use, the produced 3D computer models and 3D printed models were of a high degree of accuracy allowing for use as an effective tool for preoperative planning for patients with complex spatial anatomical relationships between cardiovascular structures or as a teaching tool for medical students.

A final limiting factor to the study was the availability of medical data. Whilst studies have been conducted which can defend the accuracy of 3D image processing based upon medical data[3,16,17,19], the use of only a single set of patient data provided no comparison between models and allowed only to demonstrate the ability of the novel image-processing technique to 3D render a single aortic aneurysm. A wider range of patient data could provide a clearer picture as to the overall capabilities of astrophysical image-processing techniques when used to render a variety of cardiovascular diseases.

## Future work

Further developments could be made using a larger sample size of data encompassing a wider variety of cardiovascular diseases in order to examine the wider capabilities of adapting and applying astrophysical image-processing techniques when examining complex cardiovascular patient anatomy. The introduction of new image-processing software packages with the capability to be programmed could also test the applicability of more complex image-processing algorithms such as the use of *noise-based detection*[20] in order to reduce the impact of false detections within the CT data using the ambient noise of surrounding pixels as a reference. Furthermore, research into the applications of astrophysical image-processing techniques within the field of cardiology using software capable of processing image data prior to thresholding, such as IDL[21], could be implemented. The introduction of this could allow for the use of initial image smoothing for obtaining large heart structures such as heart chambers and localised image enhancement[5] for improved clarity of small-scale cardiovascular features and their inclusion in complete heart models.

Future studies could delve into areas of emerging technology such as 3D bioprinting[22] utilizing astrophysical image-processing techniques in order to produce engineered tissue constructs with complex hierarchical structures, improving upon the accuracy of previously-implemented methods within the field in order to build upon previous research and bring 3D bioprinting a stage closer to clinical applications.

# Conclusions

We have detailed a step-by-step blueprint of the methodology for producing a volume rendering of a patient's heart, starting with the DICOM data and ending with the creation of a tactile 3D printout, specifically by utilizing (free) open source software, namely *Slicer 4.8.1*, *MeshLab* and *AutoDesk Netfabb*. In addition, the use of image-processing techniques inspired by astrophysics has been utilized alongside traditional medical image-processing methods in the rendering of patient-specific computer and 3D printed models. We find that it is feasible to produce realistic 3D printed cardiovascular models based on this method, and that the models produced offered an improved view of heart defects due to an improved spatial anatomical orientation. The 3D printouts showed clear views of each of the four heart chambers and major blood vessels, as well as clear tricuspid structure of the aortic valve.

The use of astrophysical image-processing techniques as inspiration for the processing techniques chosen in this paper have produced 3D models of similar quality to current models produced using medical image-processing methods:

- The use of *vignetting*, the process of creating a border around an image in order to reduce areas of ambient bright pixels, such as around a sunspot, allowed for the observation and segmentation of valve cusp intersections.

- The use of *dilation* and *boxcar smoothing* in addition to this provided an accurate approach capable of capturing valve tissue within a specified region whilst minimizing the inclusion of unwanted pixels from the volume rendering (these techniques were used to reduce the impact of hot and dark pixels on the segmented data).

Therefore, we have found that methods traditionally utilized to observe astrophysical phenomena can be adapted and applied alongside medical image-processing techniques and can be implemented to produce accurate three-dimensional renderings of patient data.

In recent years, medical model 3D printing based on CT or magnetic resonance imaging (MRI) data has been established as a complimentary imaging technique in cardiology[3,19]. The use of 3D renderings of patient data improves on traditional imaging techniques where surgeons are required to visualise a three-dimensional picture of heart defects based on a series of 2D scans by reproducing exact, real three-dimensional cardiovascular anatomy[3]. The use of 3D modelling can also improve the physician's understanding of individual patient anatomy such as in the case of valve replacement[16] or in procedural planning for the treatment of congenital heart disease[17].

# Acknowledgements

Professor James McLaughlin acknowledges STFC for IDL support as well as support via ST/L006243/1.

# References

1. Bartel T, Rivard A, Jimenez A, Mestres C.A, and Müller S. Medical Three-Dimensional Printing Opens up New Opportunities in Cardiology and Cardiac Surgery. Eur Heart J 2017; 0: 1-9
2. Nicholls M. Three-Dimensional Imaging and Printing in Cardiology. Eur Heart J 2017; 0: 230-231
3. Olejník P, Nosal M, Havran T, Furdova A, Cizmar M, Slabej M, Thurzo A, Vitovic P, Klvac M, Acel T and Masura J. Utilisation of Three-Dimensional Printed Heart Models for Operative Planning of Complex Congenital Heart Defects. Pol Heart J 2017; 0022: 1-19
4. Aschwanden M.J. Image Processing Techniques and Feature Recognition in Solar Physics. Solar Phys 2010; 262: 235-275
5. Gonzalez R.C. Woods R.E. Digital Image Processing. 3rd edn. Pearson Prentice Hall, Upper Saddle River; 2008. p38-612
6. Gurman J.B. SOHO and the Great Observatory: Proposal to the Senior Review of Sun-Solar System Connections Operating Missions. https://sohowww.nascom.nasa.gov/publications/soho_sr05_proposal.pdf (1 May 2018)
7. Phillips K J H, Chifor C and Landi E. The High-Temperature Response of the TRACE 171 Å and 195 Å Channels. The astrophysical Journal 2005; 626: p1110-1115

8. NASA. STEREO Overview. https://www.nasa.gov/mission_pages/stereo/mission/index.html (1 May 2018)
9. Grimm T. User's Guide to Rapid Prototyping Society of Manufacturing Engineers 2004: p55-58
10. Davis C, Bewsher D, Davies J and Crothers S. Working with data from the NASA STEREO Heliospheric Imager an essential guide 2008 http://www.sstd.rl.ac.uk/Stereo/Documents/HI_user_guidejuly.pdf.
11. Alizadeh M, Cote M and Albu A.B. Leaflet Free Edge Detection for the Automatic Analysis of Prosthetic Heart Valve Opening and Closing Motion Patterns from High Speed Video Recordings. Scandinavian Conference of Image Analysis 2017; 225-238
12. Meshlab. Meshlab home. http://www.meshlab.net/ (1 May 2018)
13. Thurgood J O, Morton R J and McLaughlin J A. First direct measurements of transverse waves in solar polar plumes. The Astrophysical Journal 2014; 790: L2
14. Weberg M, Morton R and McLaughlin J A. An Automated Algorithm for Identifying and Tracking Transverse Waves in Solar Images. The Astrophysical Journal 2018; 852: p57-66
15. Huotilainen E, Jaanimets R, Valášek J, Marcián P, Salmi M, Tuomi J, Mäkitie A and Wolff J. Inaccuracies in Additive Manufactured Medical Skull Models Caused by the DICOM to STL Conversion Process. Journal of Cranio-Maxillo-Facial Surgery 2014; 42: p259-265
16. Sodian R, Schmauss D, Markert M, Weber S, Nikolaou K, Haeberle S, Vogt F, Vicol C, Lueth T, Reichart B and Schmitz C. Three-Dimensional Printing Creates Models for Surgical Planning of Aortic Valve Replacement After Previous Coronary Bypass Grafting. Ann Thorac Surg 2008; 85: p2105-2109
17. Mottl-Link S, Hübler M, Kühne T, Rietdorf U, Krueger J, Schnackenburg B, DeSimone R, Berger F, Juraszek A, Meinzer H-P, Karck M, Hetzer R and Wolf I. Physical Models Aiding in Complex Congenital Heart Surgery. Ann Thorac Surg 2008; 86: p273-277
18. Valverde I, Gomez G, Gonzalez A, Suarez-Mejias C, Adsuar A, Coserria J-F, Uribe S, Gomez-Cia T and Hosseinpour A-R. Three-Dimensional Patient-Specific Cardiac Model for Surgical Planning in Nikaidoh Procedure. Cardiology in the Young 2015; 25: p698-704
19. Valverde I, Gomez G, Suarez-Mejias C, Hosseinpour A-R, Hazekamp M, Roest A, Vazquez-Jimenez J, El-Rassi I, Uribe S and Gomez-Cia T. 3D printed cardiovascular models for surgicalplanning in complex congenital heart diseases. J Cardiov Magn Reson 2015; 17(Suppl 1): 196. doi: 10.1186/1532-429X-17-S1-P196
20. Akhlaghi M and Ichikawa T. Noise-Based Detection and Segmentation of Nebulous Objects. The Astrophysical Journal 2015; 220: p1-33
21. Harris Geospatial Solutions. IDL. http://www.harrisgeospatial.com/SoftwareTechnology/IDL.aspx (1 May 2018)
22. Duan B. State-of-the-Art Review of 3D Bioprinting for Cardiovascular Tissue Engineering. Ann. Biomed Eng 2016; 45: 95-209